\documentclass[aps,prb,amssymb,showpacs,twocolumn]{revtex4}

\newcommand{\pib}{\mbox{\boldmath $\pi $}}

\newcommand{\bA}{{\bf A}}

\newcommand{\bk}{{\bf k}}

\newcommand{\bq}{{\bf q}}

\newcommand{\bK}{{\bf K}}
\newcommand{\bp}{{\bf p}}

\newcommand{\br}{{\bf r}}

\newcommand{\ef}{\epsilon_{\rm F}}
\newcommand{\vf}{v_{\rm F}}
\newcommand{\op}{\omega^{\prime}}
\newcommand{\sgn}{{\rm sgn}}
\newcommand{\xn}{\xi_n}
\newcommand{\xnp}{\xi_{n^{\prime}}}
\newcommand{\la}{\lambda}
\newcommand{\lap}{\lambda^{\prime}}
\newcommand{\np}{n_{>}}
\newcommand{\nm}{n_{<}}

\usepackage{graphics,graphicx,amsmath}

\usepackage{tabularx,float}
\usepackage{epsfig}

\usepackage{subfigure}

\usepackage{bm}
\usepackage{wasysym}

\usepackage{bm}
\usepackage[usenames]{color}
\usepackage{verbatim}

\begin{document}

\bibliographystyle{apsrev}

\date{\today}

\author{R. Rold\'{a}n, J.-N. Fuchs, and M. O. Goerbig}
\affiliation{\centerline{Laboratoire de Physique des Solides, Univ. Paris-Sud, CNRS, UMR 8502, F-91405 Orsay Cedex, France}}

\title{Collective modes of doped graphene and a standard 2DEG in a strong magnetic field:\\ linear magneto-plasmons versus magneto-excitons}

\begin{abstract}
A doped graphene layer in the integer quantum Hall regime reveals a
highly unusual particle-hole excitation spectrum, which is
calculated from the dynamical polarizability in the random phase approximation. We find that the elementary
neutral excitations in graphene in a magnetic field are unlike those
of a standard two-dimensional electron gas (2DEG): in addition to the upper-hybrid
mode, the particle-hole spectrum is reorganized in linear
magneto-plasmons that disperse roughly parallel to $\omega=\vf q$, instead of the usual horizontal (almost dispersionless) magneto-excitons. These modes could be detected in an inelastic light scattering experiment. 
\end{abstract}


\maketitle

\section{Introduction}
Particle-hole excitations in a standard two-dimensional electron gas are incoherent and form a continuum in the energy-momentum
plane.\cite{GV05,PN66} This continuum does not have a
uniform weight but features some structure that hints at possible
coherent excitations. If one includes interactions between
the electrons, coherent excitations may emerge from the continuum.
This is the case of the plasmon, which appears once long-range
Coulomb interactions are taken into account. The plasmon mode is long-lived
outside the continuum, but it is Landau damped once it penetrates
it. If instead of including interactions, one turns on a magnetic field, the
electronic energy is quenched into Landau levels (LLs), and the
particle-hole continuum becomes discrete in energy, while remaining essentially
continuous in the momentum direction. When both the magnetic field and Coulomb interactions are
considered, the continuum is reorganized in magneto-excitons --
transverse excitations that now acquire a dispersion\cite{KH84} --
and an upper-hybrid mode\cite{CQ74} exists
outside what used to be the continuum.


Graphene, a recently discovered carbon material, is attracting a lot of interest due to its unique electronic properties (see Ref. \onlinecite{CG07}). Electrons in graphene may be viewed as a particular form of the 2DEG. However, due to its underlying
triangular lattice with a two-atom basis, the electrons of graphene are
described by a massless Dirac equation instead of the usual
effective-mass Schr\"odinger equation. The screening properties of graphene are different from that of the 2DEG, as it may be seen from the polarization and dielectric functions in the two cases.\cite{S86,GGV94,WSSG06,HS07,SNC08} Furthermore, the single-particle spectral function reveals the particular chiral properties of the Dirac-like quasiparticles.\cite{PM08,HS08} Another salient physical consequence is a peculiar integer quantum Hall effect.\cite{NF05,ZTSK05}

In this paper, we show that collective excitations of graphene in a
strong magnetic field are much unlike that of the standard 2DEG [from now on, we use the term 2DEG to denote a standard 2DEG with a parabolic band dispersion, in contrast to graphene]. The
magnetic field also reorganizes the particle-hole continuum but
instead of revealing horizontal lines, diagonal lines emerge. When
Coulomb interactions are included within the random-phase approximation (RPA), the diagonal excitations acquire coherence. We shall refer to them as \emph{linear
magneto-plasmons}  in order to distinguish them from the
\emph{upper-hybrid mode} -- which is also present and which may be considered as the plasmon mode modified by the
magnetic field -- and from the usual
(horizontal) \emph{magneto-excitons}, which are blurred in
graphene. Several recent theoretical works considered
collective excitations in graphene in a magnetic field,
concentrating either on magneto-excitons\cite{IJFB07,BM08,S07} or on the plasmon
mode.\cite{TS08,BGL08} 
Here, we show that concentrating on horizontal magneto-excitons, in spite of its success in the 2DEG,\cite{KH84}
is not sufficient to reveal the complete structure of the particle-hole excitation spectrum (PHES) in 
graphene, which is dominated by linear magneto-plasmons that disperse roughly parallel to $\omega=v_Fq$.

The paper is organized as follows. In Sec. II we calculate the polarization function of a 2DEG and that of doped graphene in the integer quantum Hall regime. In Sec. III we discuss the main features of the non-interacting PHES of graphene and a 2DEG, and the different nature of the collective modes in each case, once electron-electron interactions are taken into account. The conclusions are summarized in Sec. IV. 

\section{Polarization function}

The polarization operator $\Pi(\bq,\omega)$ may be viewed as the
particle-hole pair propagator, the poles of which yield the dispersion
relation and damping of the collective excitations. Its imaginary
part is related to the dynamical structure factor by
$$S(\bq,\omega)=-\frac{1}{\pi}{\rm Im}\, \Pi (\bq,\omega), $$ which plays the
role of a spectral function for the particle-hole propagator. The PHES is defined as the
$(\omega,q)$ region of non-zero spectral weight $S(\bq,\omega)\neq
0$. Peaks in the spectral density are interpreted as collective
excitations; their dispersion relation and damping may be extracted
from the position and the width of the peaks (see e. g. \S 3.2.7. in Ref. \onlinecite{GV05}).

The Fourier-transformed polarizability for non-interacting electrons
(we use a unit system with $\hbar\equiv 1$),
\begin{equation}\label{Pi0}
\Pi^0(\bq,\omega)\!=\!\!\int \!\!\frac{d\op d\bk}{i(2\pi)^3}\,{\rm
Tr}\!\!\left[ G^0(\bk,\op)G^0(\bk+\bq,\op+\omega)\right ],
\end{equation}
is given in terms of the free single-particle Green's functions $G^0(\bq,\omega)$. Whereas, as we
review below, the single-particle Green's functions for the conventional 2DEG are simple scalar
functions, those of graphene are $2\times 2$ matrices due to the two sublattices $A$ and $B$.
The trace (Tr) in Eq. (\ref{Pi0}) is, thus, relevant only in the case of graphene.

\subsection{Polarizability in the 2DEG}

The Hamiltonian for electrons in the conventional 2DEG is simply that of free electrons with a band mass $m_b$, ${\cal H}=\pib^2/2m_b$, where the gauge-invariant momentum operators,
$\pib=\bp+e\bA(\br)$ takes into account the coupling to the magnetic field ${\bf B}=\nabla\times \bA$, where $\bA$ is the vector potential.
The operators $\pib$ may be expressed in terms of the usual ladder operators with the help of 
$$a=\frac{l_B}{\sqrt{2}}(\pi_x-i\pi_y), \qquad a^{\dagger}=\frac{l_B}{\sqrt{2}}(\pi_x+i\pi_y),$$
which satisfy the commutation relations $[a,a^{\dagger}]=1$, 
and $l_B=1/\sqrt{eB}$ is the magnetic length. One obtains the usual eigenstates,
$|n,m\rangle$, and the eigenvalue equation ${\cal H}|n,m\rangle=\omega_c(n+1/2)|n,m\rangle$, where
$\omega_c=1/m_b l_B^2=eB/m_b$ is the cyclotron frequency. The result is the usual LL spectrum where the
levels are labeled by the quantum number $n$, the eigenvalue of $a^{\dagger}a$.
The additional quantum number $m$
is associated with the guiding center operator, and varies between $0$ and $N_B-1$,
where $N_B={\cal A}/2\pi l_B^2$, in terms of the sample area $\cal A$. 

With the help of the eigenstates $|n,m\rangle$, one may express the Green's functions $G^0(\bq,\omega)$
for non-interacting electrons in Fourier space, as
\begin{equation}\label{Green:2DEG}
 G^0(\bq,\omega)=\sum_{n,m}\frac{\langle \bq|n,m\rangle\langle n,m|\br=0\rangle}{\omega-\xi_n+i\delta\, \sgn(\xi_n)}\ ,
\end{equation}
where $\xi_n\equiv \omega_c(n+1/2)-\ef$ is the energy of the $n$-th LL with respect to the Fermi energy $\ef$, which we choose to lie between
two LLs of the conduction band (integer quantum Hall regime), and 
$\delta$ accounts for disorder-induced 
level broadening ($\delta \to 0^+$ in the clean limit).

Eq. (\ref{Green:2DEG}) allows us to calculate the polarization function (\ref{Pi0}) for the 2DEG,
\begin{equation}\nonumber
\Pi^0(\bq,\omega)=\sum_{n=0}^{N_F}\sum_{n'=N_F+1}^{\infty}\frac{F_{nn'}(\bq)}{(n-n')\omega_c+\omega+i\delta}+(\omega^+\rightarrow\omega^-)\ ,
\end{equation}
where $N_F$ is the index of the last occupied LL, which is fixed by the filling factor, and $\omega^+\rightarrow\omega^-$ indicates the replacement $\omega+i\delta\rightarrow-\omega-i\delta$. The 
form factors for the 2DEG read\cite{GV05}
\begin{equation}\label{form:2DEG}
 F_{nn'}(\bq)\!=\!e^{-l_B^2q^2/2}\!\left( \frac{l_B^2q^2}{2}\right)^{\np-\nm}\!\!\frac{\nm !}{\np!}\!
\left [L_{\nm}^{\np-\nm}\left (\frac{l_B^2q^2}{2}\right )\right ]^2,
\end{equation}
with $\np\equiv \max\{n,n'\}$ and $\nm\equiv \min \{n,n'\}$, and $L_n^m$ are associated Laguerre polynomials.

\subsection{Polarizability in graphene}

For graphene in a magnetic field, the electronic Hamiltonian may be written as\cite{CG07}
\begin{equation}\label{Ham:Dir}
 {\cal H}=\frac{\sqrt{2}\vf}{l_B}\left(\begin{array}{cc}
 0 & a\\ a^{\dagger} & 0
\end{array}
\right)\ ,
\end{equation}
where $\vf=3ta_{cc}/2$ is the Fermi velocity, in terms of the the nearest-neighbor
hopping integral $t\simeq 3$eV, and the carbon-carbon distance $a_{cc}\simeq 1.4{\rm\AA}$.
Strictly speaking, Eq. (\ref{Ham:Dir}) is only valid on one of the valleys, namely $K$, that
of the valley $K'$ being $-{\cal H}^*$. However, we concentrate, here, only on processes that do 
not couple the two different valleys, such that a discussion of the Hamiltonian (\ref{Ham:Dir})
is sufficient, and the twofold valley degeneracy may be accounted for by a simple factor of 2.

The Hamiltonian (\ref{Ham:Dir}) yields the relativistic LL spectrum 
\begin{equation}
\lambda \epsilon_n=\lambda\frac{\vf}{l_B}\sqrt{2n},
\nonumber
\end{equation}
where $\lambda$ denotes the band index, $\lambda=+$ for the conduction and $\lambda=-$ for the
valence band. The associated eigenstates are the 2-spinors 
\begin{eqnarray}
\nonumber
\psi_{\lambda,n,m} = \frac{1}{\sqrt{2}}\left(\begin{array}{c}
|n-1,m\rangle \\ \lambda |n,m\rangle \end{array}\right) \qquad &&{\rm for~} n\geq 1 \ ,\\
\nonumber
\psi_{n=0,m} = \left(\begin{array}{c}
0 \\ |n=0,m\rangle \end{array}\right) \qquad &&{\rm for~} n=0\ .
\end{eqnarray}
where $m=0,1,...,N_B-1$, and $|n,m\rangle$ are the
corresponding eigenstates of the Hamiltonian with quadratic
dispersion, introduced in the previous subsection. Due to the 2-spinor structure of the wavefunctions in graphene, the associated single-particle 
Green's functions are $2\times 2$ matrices,
\begin{equation}
G_{\zeta;\alpha\alpha^{\prime}}^0(\bk ,\omega)=
\sum_{\la}\sum_{n}\frac{f_{\zeta,\alpha\alpha^{\prime};\la n}(\bk +\zeta\bK) }{\omega -\la\xi_n +i\delta \sgn(\la\xi_n)},\\
\end{equation}
with $\alpha=A(B)$ for electrons on the $A(B)$ sublattice,
$\zeta=+(-)$ for electrons in the $K$($K'$)-valley, and
$\bk$ is the electron momentum measured from the Dirac
points, $\zeta\bK=\zeta 4\pi/3\sqrt{3}a{\bf u}_x$. In the
denominator, we have defined $\la\xi_n\equiv\la\epsilon_n-\ef$, the energy difference between the
level $\la\epsilon_n$ and the Fermi energy $\ef$. Furthermore, we neglect Zeeman and possible valley splittings.
The matrix $f_{\la n}(\bq)$ for the $K$-valley is
\begin{widetext}
\begin{equation}
f_{K;\lambda n}(\bq)=\sum_m \left (
\begin{array}{cc}
1_n^{*2}\langle \bq  |n-1,m\rangle \langle n-1,m|\br =0\rangle & -i\lambda1_n^*2_n^*\langle \bq  |n-1,m\rangle \langle n,m|\br =0\rangle \\
i\lambda1_n^*2_n^*\langle \bq  |n,m\rangle \langle n-1,m|\br
=0\rangle & 2_n^{*2}\langle \bq  |n,m\rangle \langle n,m|\br
=0\rangle
\end{array}
\right )\, ,
\end{equation}
\end{widetext}
where we have introduced the simplified notation
$1^*_n=\sqrt{(1-\delta_{n,0})/2}$,
$2^*_n=\sqrt{(1+\delta_{n,0})/2}$. 
The bare polarization bubble is
\begin{equation}\label{EqPi0}
\Pi^0(\bq,\omega)=\sum_{n=1}^{N_F}\Pi_{n}^{\la_F}(\bq,\omega)+\Pi^{vac}(\bq,\omega)\,
,
\end{equation}
where the index $N_F$ of the last occupied LL is now related
to the filling factor $\nu$ by $\nu=4N_F+2$ due to the fourfold spin and
valley degeneracy. There is a non-zero vacuum polarization,
$\Pi^{vac}(\bq,\omega)\equiv
-\sum_{n=1}^{N_c}\Pi_{n}^{\la=1}(\bq,\omega)$, the vacuum
corresponding to undoped graphene. Here, $N_c$ is an ultra-violet cutoff
\footnote{The continuum approximation is valid up to an energy $\sim
t$. This leads to a cutoff in the Landau level index $N_c\sim
10^4/B[T]$. As the LL separation in graphene decreases with $n$, it
is possible to have {\it semi-quantitative} results from smaller
values of $N_c$. Here, for numerical calculations, we typically take
$N_c\approx 70$. There are no noticeable differences for larger values of $N_c$.} and
\begin{eqnarray}
\Pi_{n}^{\la}(\bq,\omega)&=&\sum_{\lap}\sum_{n'=0}^{n-1}\Pi_{nn'}^{\la\lap}(\bq,\omega)\nonumber\\
&+&\sum_{\lap}\sum_{n'=n+1}^{N_c}\Pi_{nn'}^{\la\lap}(\bq,\omega)+\Pi_{nn}^{\la
-\la}(\bq,\omega).
\end{eqnarray}
In the previous expressions, we have used
\begin{equation}
\Pi_{nn'}^{\la\lap}(\bq,\omega)\equiv\frac{{\cal
\overline{F}}_{nn^{\prime}}^{\lambda\lambda^{\prime}}(\bq)}{\lambda\xn-\lambda^{\prime}\xnp
+\omega+i\delta }+(\omega^+\rightarrow\omega^-)\, 
,
\end{equation}
where the form factor is 
\begin{widetext}
\begin{equation}\label{FF}
{\cal
\overline{F}}_{nn^{\prime}}^{\lambda\lambda^{\prime}}(\bq)=e^{-l_B^2q^2/2}\left
( \frac{l_B^2q^2}{2}\right)^{\np-\nm}\left \{
\lambda 1_n^{*}1_{n'}^{*}\sqrt{\frac{(\nm-1)!}{(\np-1)!}}\,L_{\nm-1}^{\np-\nm}\left (\frac{l_B^2q^2}{2}\right )
+\lambda'2_n^{*}2_{n'}^{*}\sqrt{\frac{\nm!}{\np!}}\,L_{\nm}^{\np-\nm}\left (\frac{l_B^2q^2}{2}\right )
\right \}^2.
\end{equation}
\end{widetext} 
Notice that Eq. (\ref{FF}) agrees with the form factor obtained in Ref. \onlinecite{S07}, but not with those obtained in Ref. \onlinecite{TS08,BGL08}.
Because of the presence of nodes in the LL wavefunctions (zeros of the Laguerre polynomials), ${\cal
\overline{F}}_{nn^{\prime}}^{\lambda\lambda^{\prime}}(\bq)$ contains 
zeros and, therefore, so does ${\rm Im}\,\Pi^0$ as a function of $q$ at fixed $\omega$. These zeros will play an important role when discussing the structure of the PHES in Sec. III.C. Note 
that $\Pi_{n}^{\la}(\bq,\omega)=-\Pi_{n}^{-\la}(\bq,\omega)$ and
that the $n=0$ LL does not contribute to the polarization.

\section{Results}

\begin{figure*}[t]
  \centering
  \subfigure[]{\label{ImPi0DP}\includegraphics[width=0.36\textwidth]{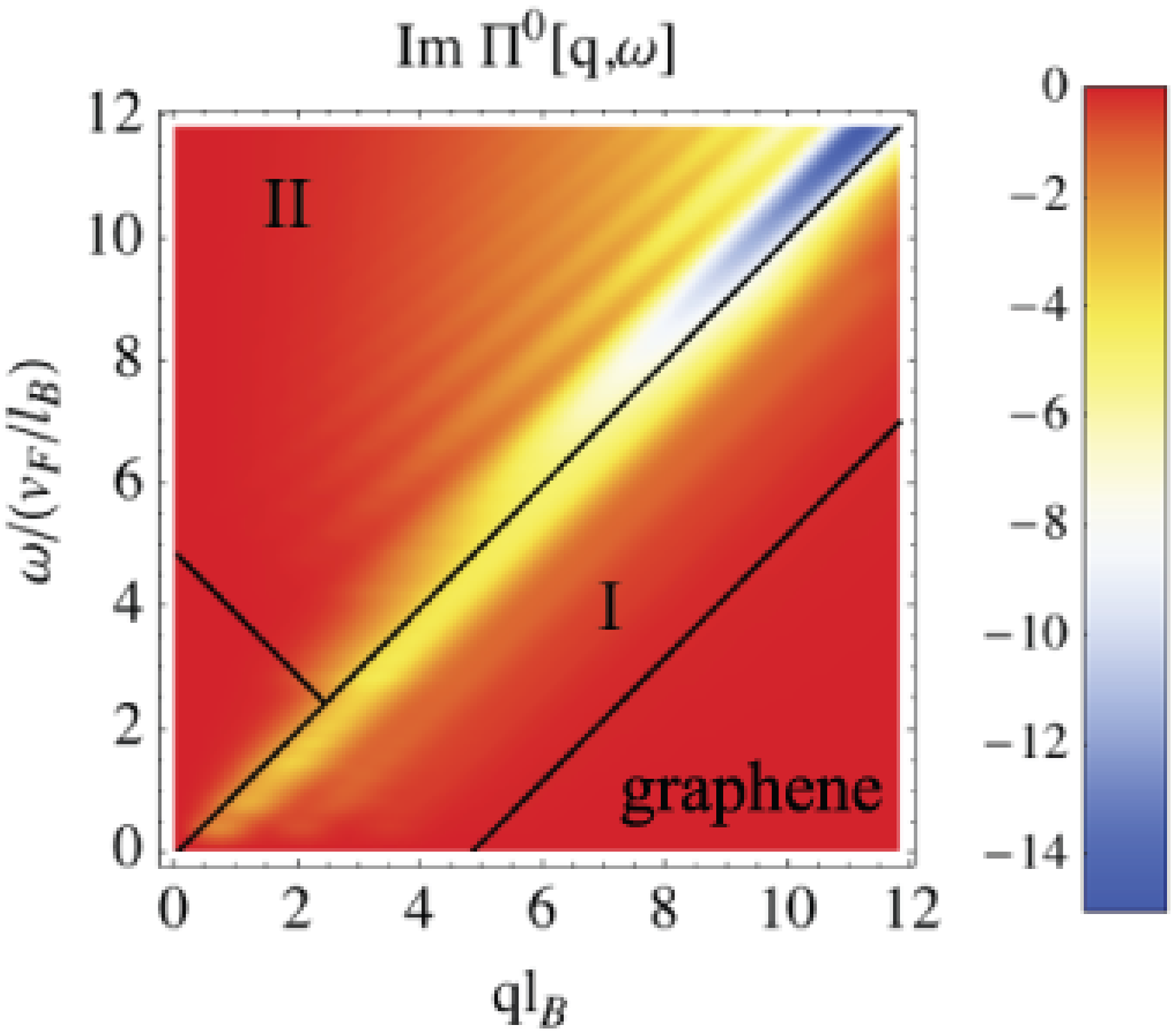}}
  \subfigure[]{\label{ImPi0DP2DEG}\includegraphics[width=0.36\textwidth]{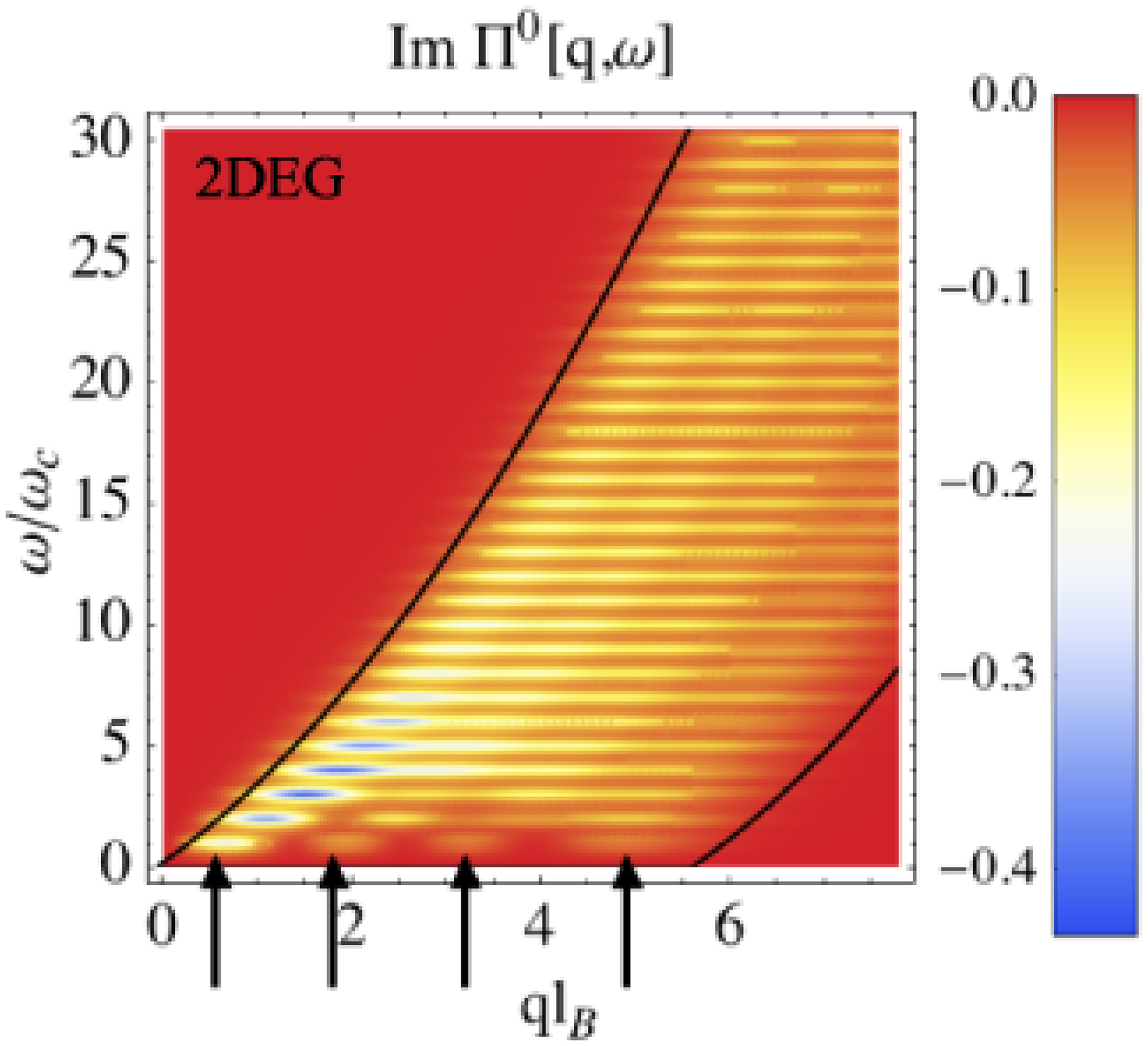}}
  \subfigure[]{\label{ImPiRPADP}\includegraphics[width=0.36\textwidth]{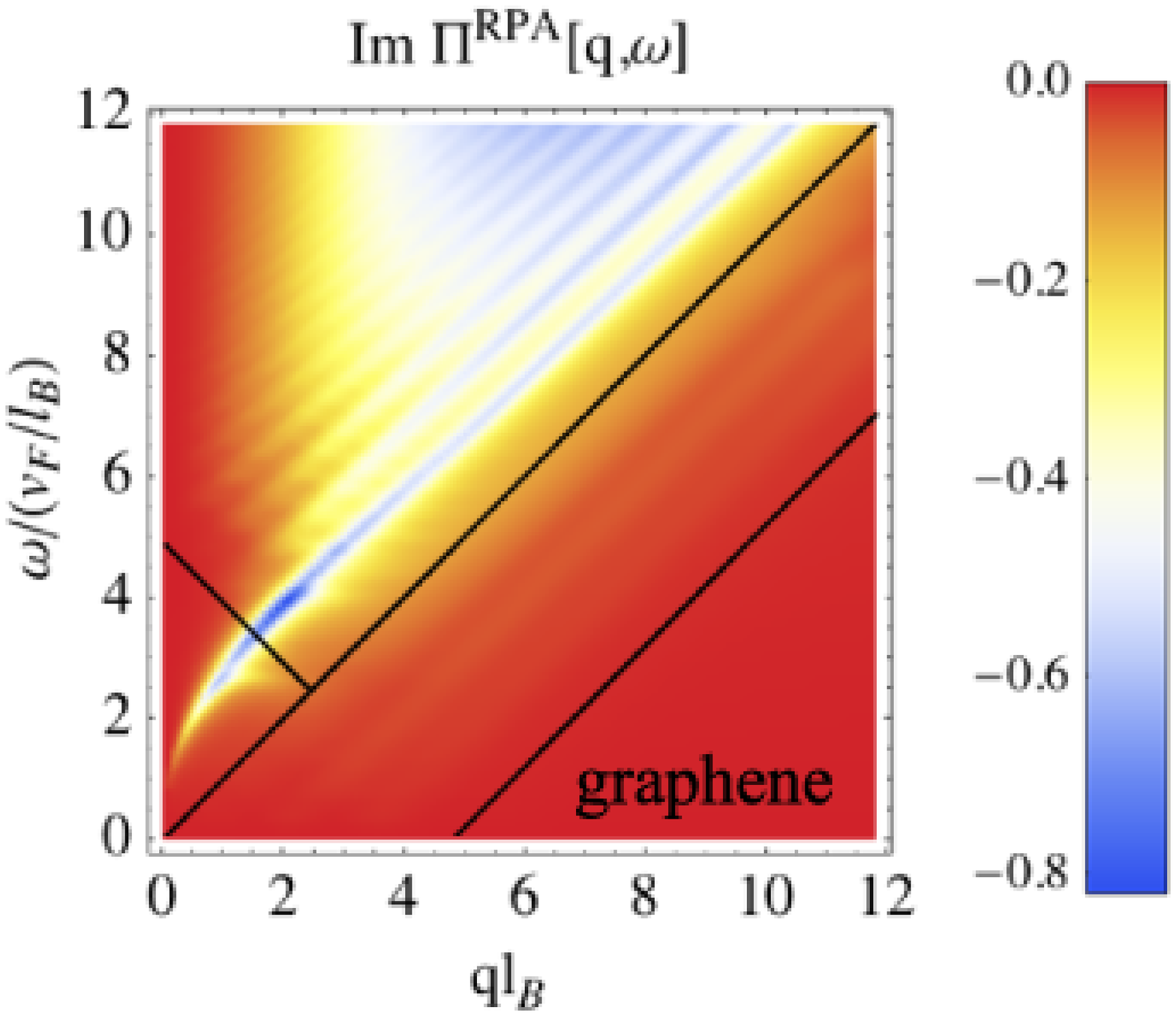}}
   \subfigure[]{\label{ImPiRPADP2DEG}\includegraphics[width=0.36\textwidth]{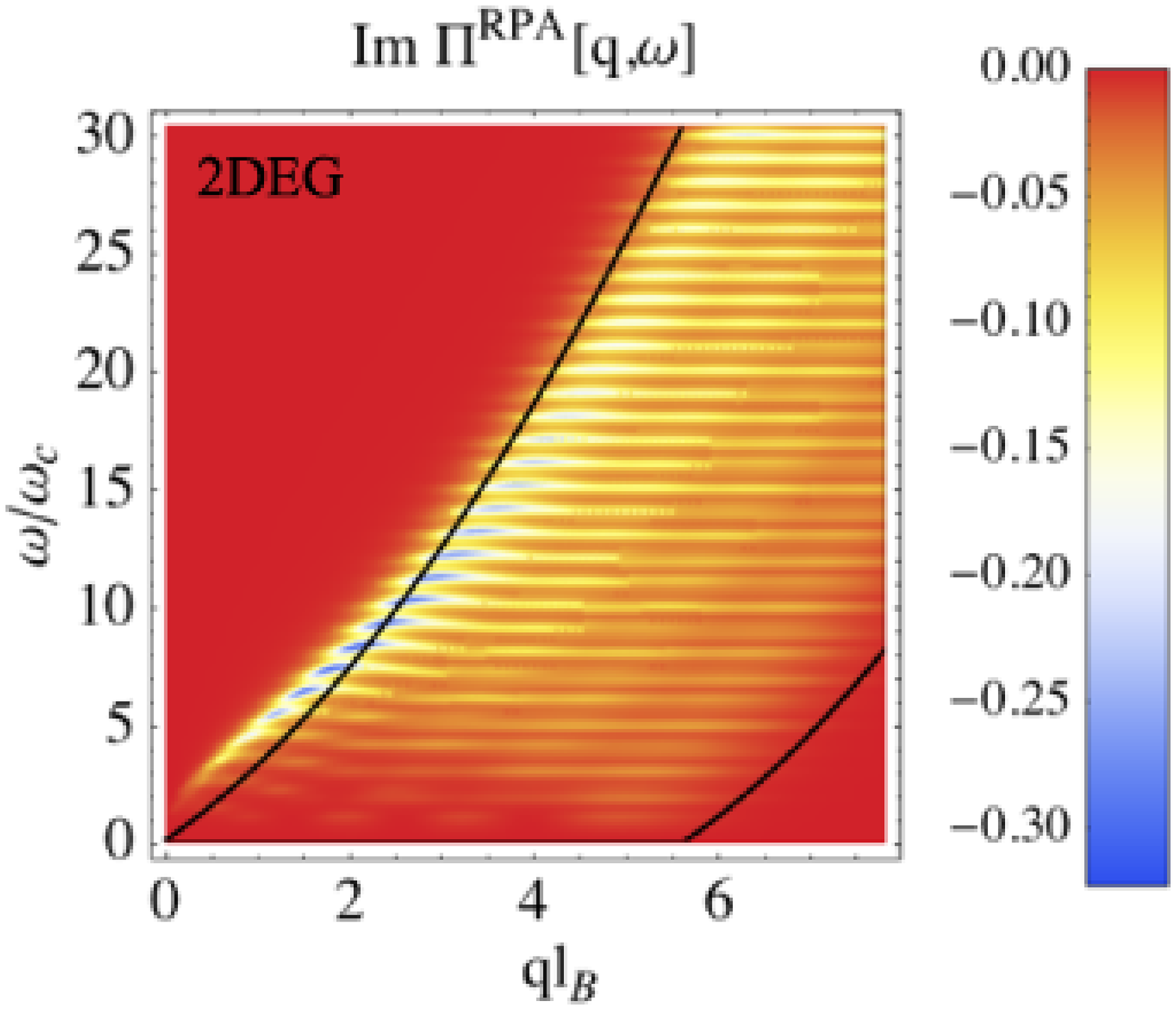}}
  \caption{\footnotesize (Color online) Density plot of the spectral function ${\rm Im}\,\Pi(\bq,\omega)$ for doping $N_F=3$, ultraviolet cutoff
 $N_c=70$ and disorder broadening $\delta=0.2\vf/l_B$ for graphene [panels (a) and (c)] and $\delta=0.2\omega_c$ for a standard 2DEG [(b) and (d)]: in the absence of interactions [(a) and (b)]
  and including interactions in the RPA [(c) with $r_s\simeq 1$ and (d) with $r_s\simeq 3$]. The solid lines indicate the zero field limits of the PHES and arrows point to the {\it islands} discussed in the text. In panel (a), label I (II) indicates the intra- (inter-) band region of the PHES in graphene.}
  \label{PiDinamic}
\end{figure*}

The dynamical polarization function
of graphene, $\Pi^0_{graph}$, is now compared to that of the 2DEG, $\Pi^0_{2DEG}$.
\cite{GV05} For $B=0$, $\Pi^0_{graph}$ has been calculated first in the context of intercalated graphite\cite{S86} and later for
doped graphene,\cite{HS07,WSSG06} resulting in a particle-hole {\it
continuum} different from that of a 2DEG. The edges of the zero-field PHES for
graphene and a 2DEG are drawn as
black lines in Fig. \ref{PiDinamic}. While the PHES for a 2DEG
is made of a single region limited by linear-parabolic boundaries
[see Fig. \ref{PiDinamic} (b)],
the PHES in graphene has linear edges and contains two different
regions because  
of intra- (I) and inter-band (II)
processes [see Fig. \ref{ImPi0DP}]. The spectral weight in graphene is concentrated around the diagonal $\omega=v_Fq$ due to the chirality factor $[1+\lambda\lambda' \cos
\theta]/2$,\cite{PM08} which is the $B=0$ equivalent of Eq. (\ref{FF}).

\subsection{Particle-hole excitation spectrum for non-interacting electrons}

In the presence of a magnetic field, the largest contribution to the
polarization also comes from the region around $\omega=\vf q$ [see Fig.
\ref{PiDinamic} (a)]. In addition, ${\rm
Im}\,\Pi_{graph}^0(\bq,\omega)$ is finite not only close to the diagonal, but also along regions above and below
$\omega=v_Fq$ [see the diagonal yellow stripes in Fig. \ref{ImPi0DP}]. For comparison, in Fig. \ref{ImPi0DP2DEG}, we show the corresponding
density plot\cite{GV05} of $\rm{Im }\,\Pi^0_{2DEG}(\bq,\omega)$ 
calculated for the same Fermi momentum $k_F \leftrightarrow
\sqrt{2N_F+1}/l_B$.

These results reveal the main differences between the PHES of
graphene and that of a 2DEG in the presence of a magnetic field. 
While the PHES for a 2DEG [Fig. \ref{PiDinamic}(b)] features a set of well-defined {\it horizontal} lines,
with a slight modulation parallel to the boundaries of the particle-hole continuum, that of graphene is
remarkably different. Indeed, its PHES [Fig. \ref{PiDinamic}(a)] is {\it dominated} by this modulation,
visible as bright regions in the diagonal, whereas the horizontal lines are now hardly visible, as a consequence of 
the non-equidistant LL spacing and the presence of a finite disorder broadening $\delta$, as discussed in more detail below [see Sec. \ref{Sec:Discuss}]. 
The PHES is, thus, discretized into 
{\it diagonal} lines (almost parallel to $\omega=\vf q$) in the $(q,\omega)$-plane.
For Fig. \ref{PiDinamic}, we have chosen $\delta =0.2\vf/l_B$, which
is a realistic value for the currently realized graphene samples.\cite{A07} Lower values of $\delta$ lead to a clearer definition of the horizontal lines [see Fig. \ref{ImPi0zoom_delt0_05} for a zoom of the low energy region of the PHES in a cleaner system].

\subsection{Particle-hole excitation spectrum for interacting electrons (RPA)}

\begin{figure*}[t]
  \centering
   \subfigure[]{\label{ImPi0zoom_delt0_05}\includegraphics[width=0.36\textwidth]{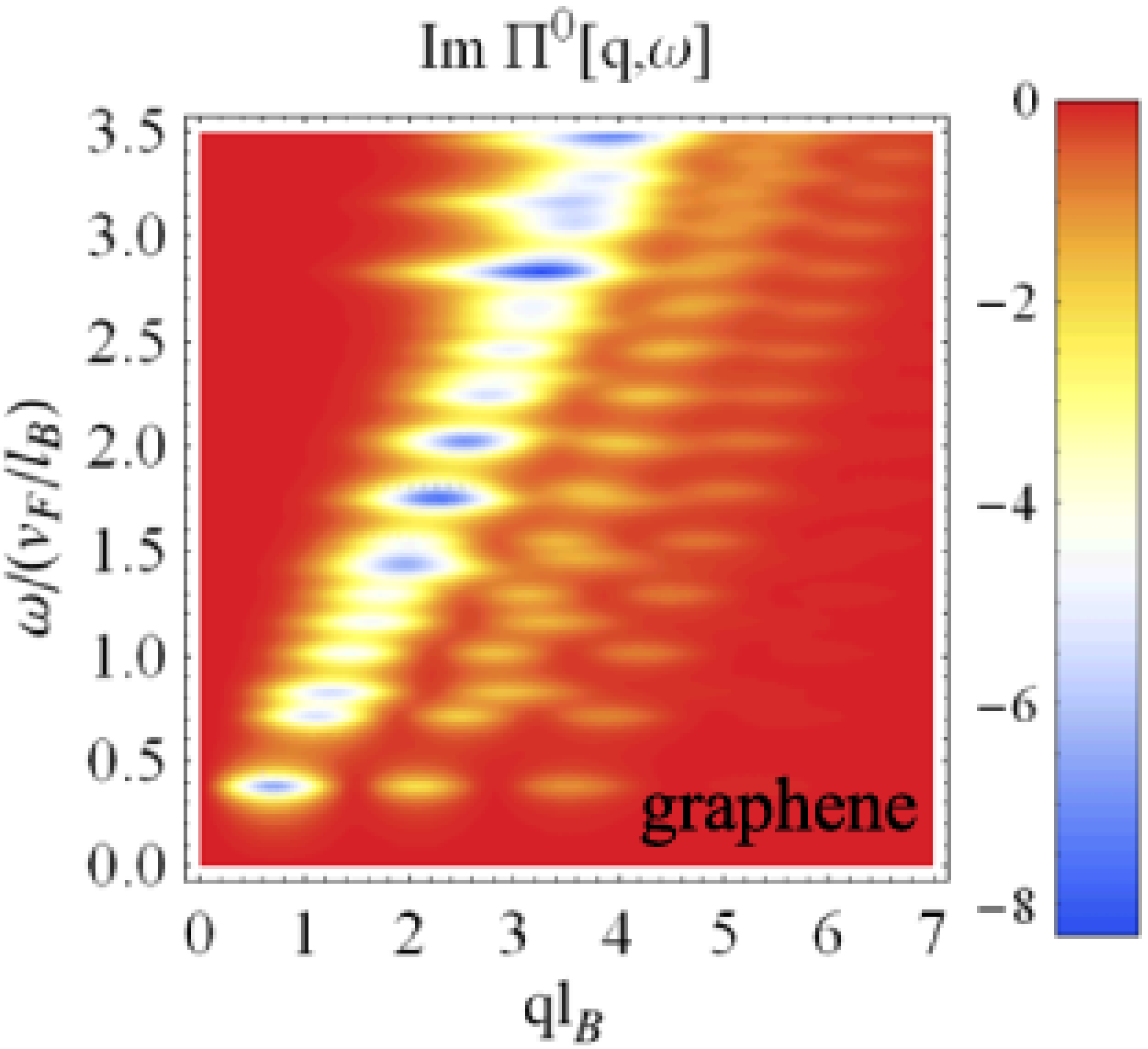}}
  \subfigure[]{\label{ImPi02DEGzoom_delt0_05}\includegraphics[width=0.36\textwidth]{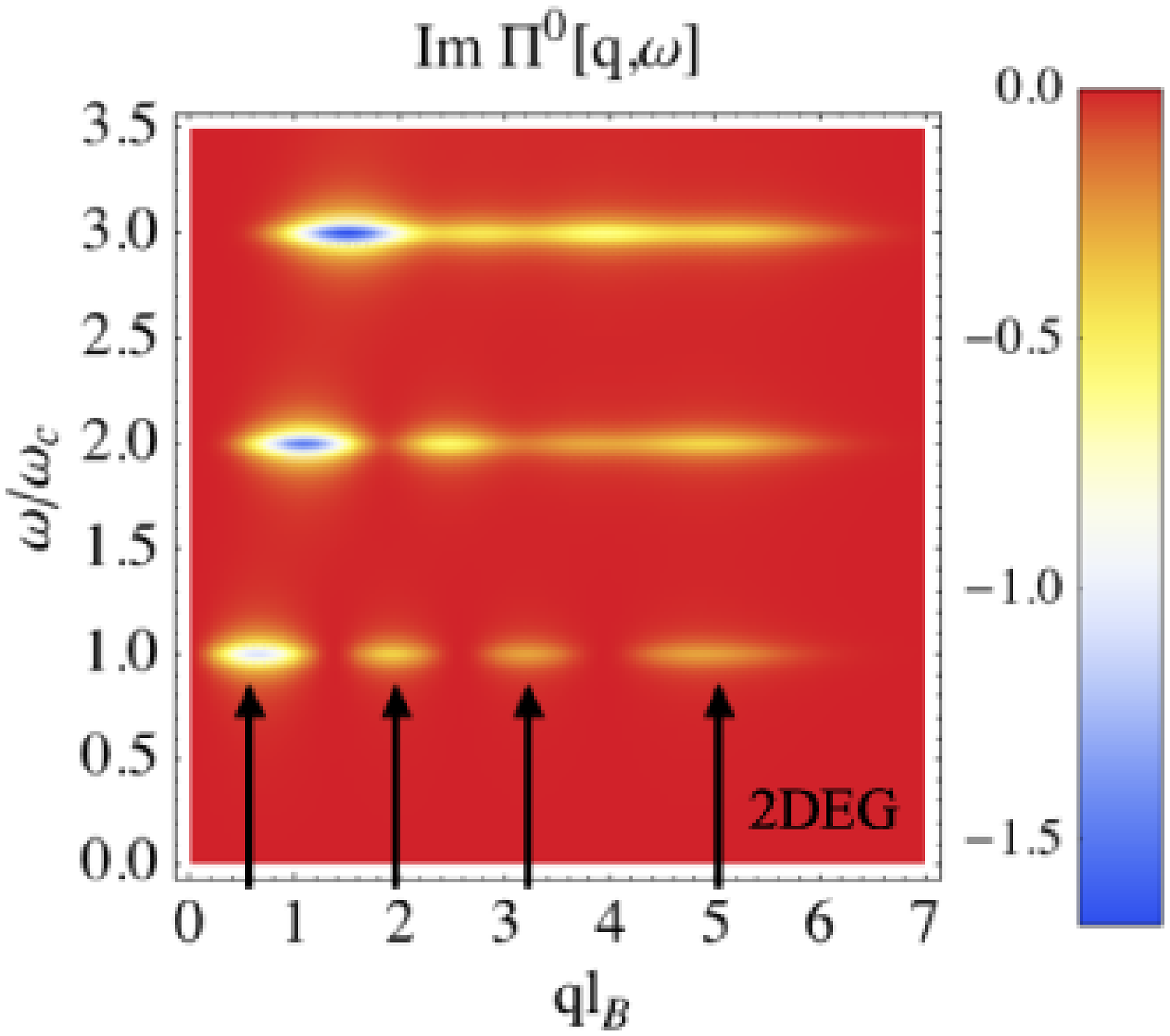}}
  \subfigure[]{\label{ImPiRPAzoom_delt0_05}\includegraphics[width=0.36\textwidth]{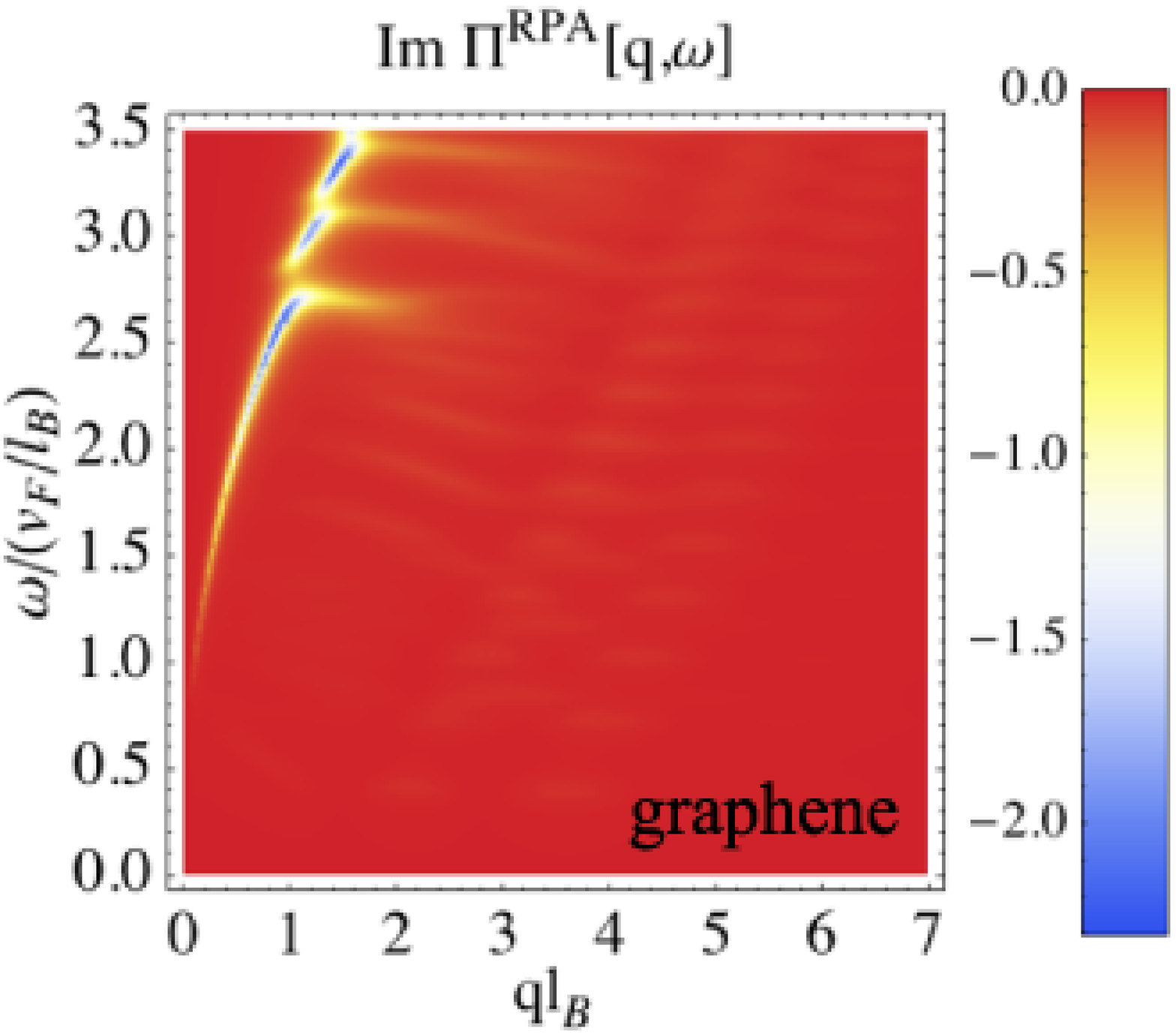}}
   \subfigure[]{\label{ImPiRPA2DEGzoom_delt0_05}\includegraphics[width=0.36\textwidth]{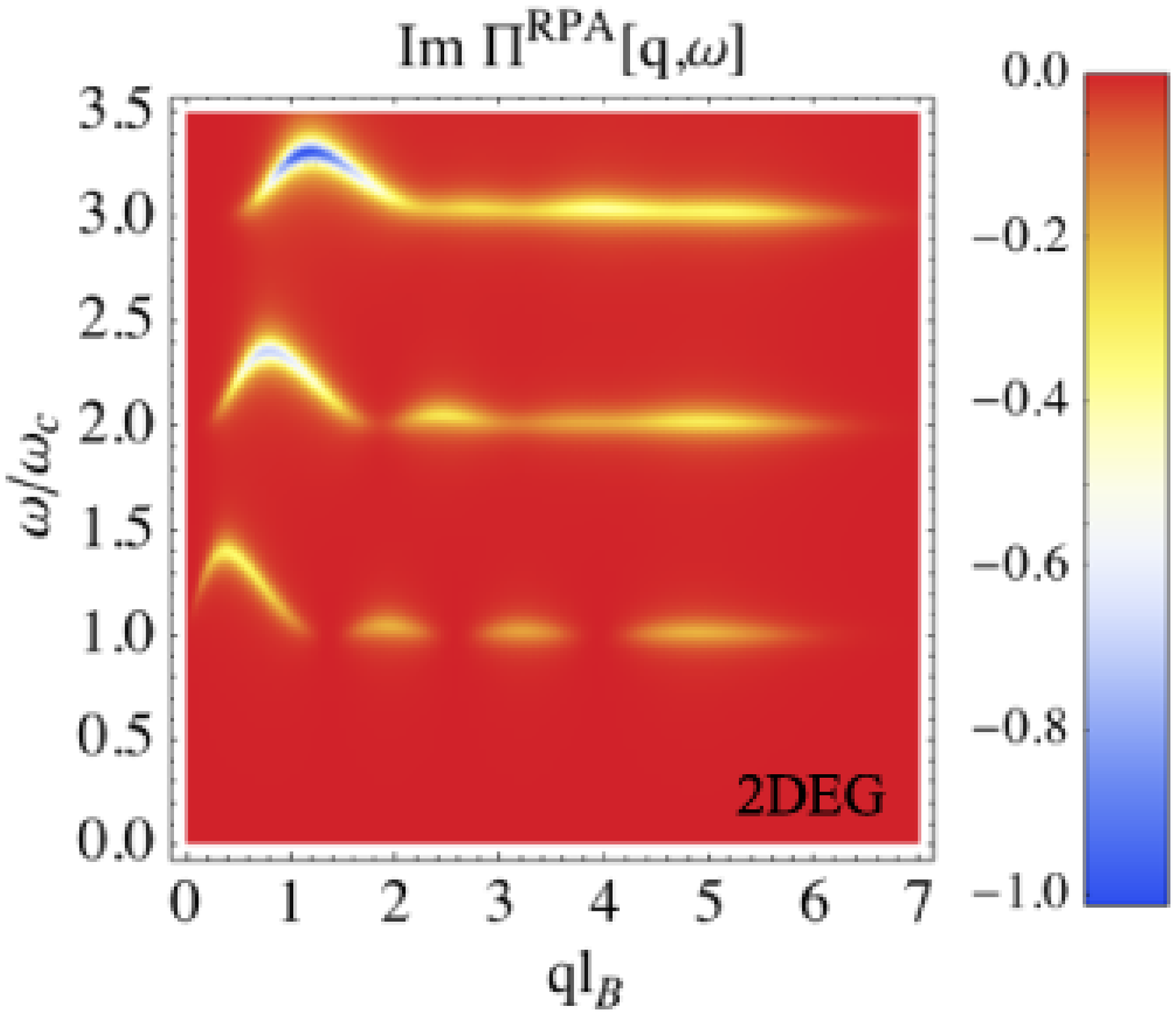}}
  \caption{\footnotesize (Color online) Density plot of ${\rm Im}\,\Pi(\bq,\omega)$ for the low energy region, for doping $N_F=3$ and ultraviolet cutoff $N_c=70$. Panels (a) and (c) correspond to graphene, whereas (b) and (d) correspond to a 2DEG. The disorder broadening $\delta=0.05$ in units of $\vf/l_B$ and $\omega_c$, respectively, and the interaction strength $r_s\simeq 1$ in panels (c) and (d). Notice the different energy scale with respect to Fig. \ref{PiDinamic}. Arrows indicate {\it islands} as discussed in the text.}
  \label{PHESzoom}
\end{figure*}

Electron-electron interactions yield coherent modes that emerge, both in the 2DEG and in graphene, from the regions of highest
weight in the PHES. Within the RPA, the
renormalized polarization function is given by 
$$
\Pi^{RPA}(\bq,\omega)=\frac{\Pi^0(\bq,\omega)}{1-v(\bq)\Pi^0(\bq,\omega)},$$ 
where
$v(\bq)=2\pi e^2/\varepsilon_b |\bq|$ is the unscreened
2D Coulomb potential, $\varepsilon_b$ the
dielectric constant, and inter-valley processes, which are suppressed in $a/l_B$, are neglected.\cite{GMD06} The results are shown in Fig. \ref{ImPiRPADP}, \ref{ImPiRPAzoom_delt0_05}
for graphene (with $r_s\equiv e^2/\varepsilon_b v_F \approx
1$) and in Fig. \ref{ImPiRPADP2DEG}, \ref{ImPiRPA2DEGzoom_delt0_05} for a 2DEG (with $r_s\equiv 2
m_b e^2/ \varepsilon_b k_F\approx 1$ or 3). 
In the 2DEG, Coulomb interactions lead to the appearance of dispersive magneto-excitons [see Fig. \ref{ImPiRPA2DEGzoom_delt0_05}], whereas in graphene, the diagonal lines of the non-interacting PHES [Fig. \ref{PiDinamic}(a)] become coherent collective modes: 
linear magneto-plasmons that are now clearly visible as peaks in the
spectral function. They disperse roughly parallel to $\omega=\vf q$ and are more pronounced in the inter-band region of the PHES [see Fig. \ref{PiDinamic}(c)]. Notice that even in cleaner samples [$\delta=0.05\vf/l_B$, Fig. \ref{ImPiRPAzoom_delt0_05}],
the horizontal structure due to the LL quantization is highly suppressed in ${\rm Im}\,\Pi^{RPA}(\bq,\omega)$.
Furthermore, there has been a transfer of spectral weight
from the long-wavelength region of the non-interacting PHES to the
upper-hybrid mode that starts dispersing in the gapped region of the
spectrum [see Fig. \ref{ImPiRPAzoom_delt0_05}]. In a 2DEG, this upper-hybrid mode\cite{CQ74} can be seen as the plasmon mode modified by the magnetic field. It is a
plasmon-cyclotron collective mode that has a dispersion relation
$\omega=[\omega_c^2+\omega_p^2]^{1/2}$, where
$\omega_p \approx \sqrt{\ef e^2 q/\varepsilon_b}$ in the long-wavelenght limit. As it is the most intense mode, Its square-root behavior is clearly
visible in Fig. \ref{PiDinamic}(d) [in Fig. \ref{ImPiRPA2DEGzoom_delt0_05} it only appears as a maximum in the magneto-excitons dispersion relation] and also in Fig. \ref{ImPiRPADP}, \ref{ImPiRPAzoom_delt0_05} and \ref{ImPiRPA_NF6_delt0_2}, (d) for graphene.

\subsection{Discussion}\label{Sec:Discuss}

\begin{figure*}[t]
  \centering
   \subfigure[]{\label{ImPiRPA_NF1_delt0_2}\includegraphics[width=0.36\textwidth]{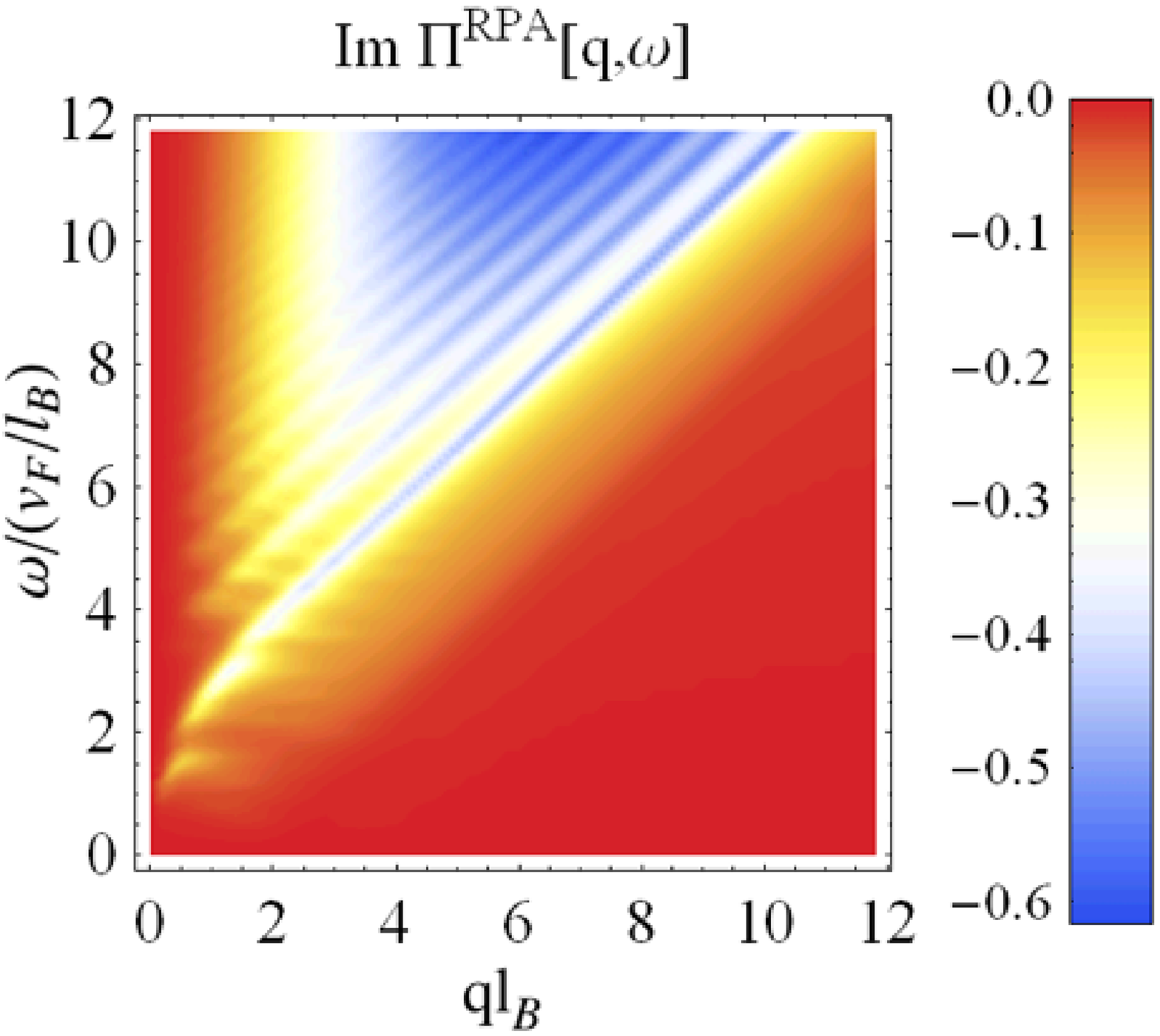}}
  \subfigure[]{\label{ImPiRPA_NF6_delt0_2}\includegraphics[width=0.36\textwidth]{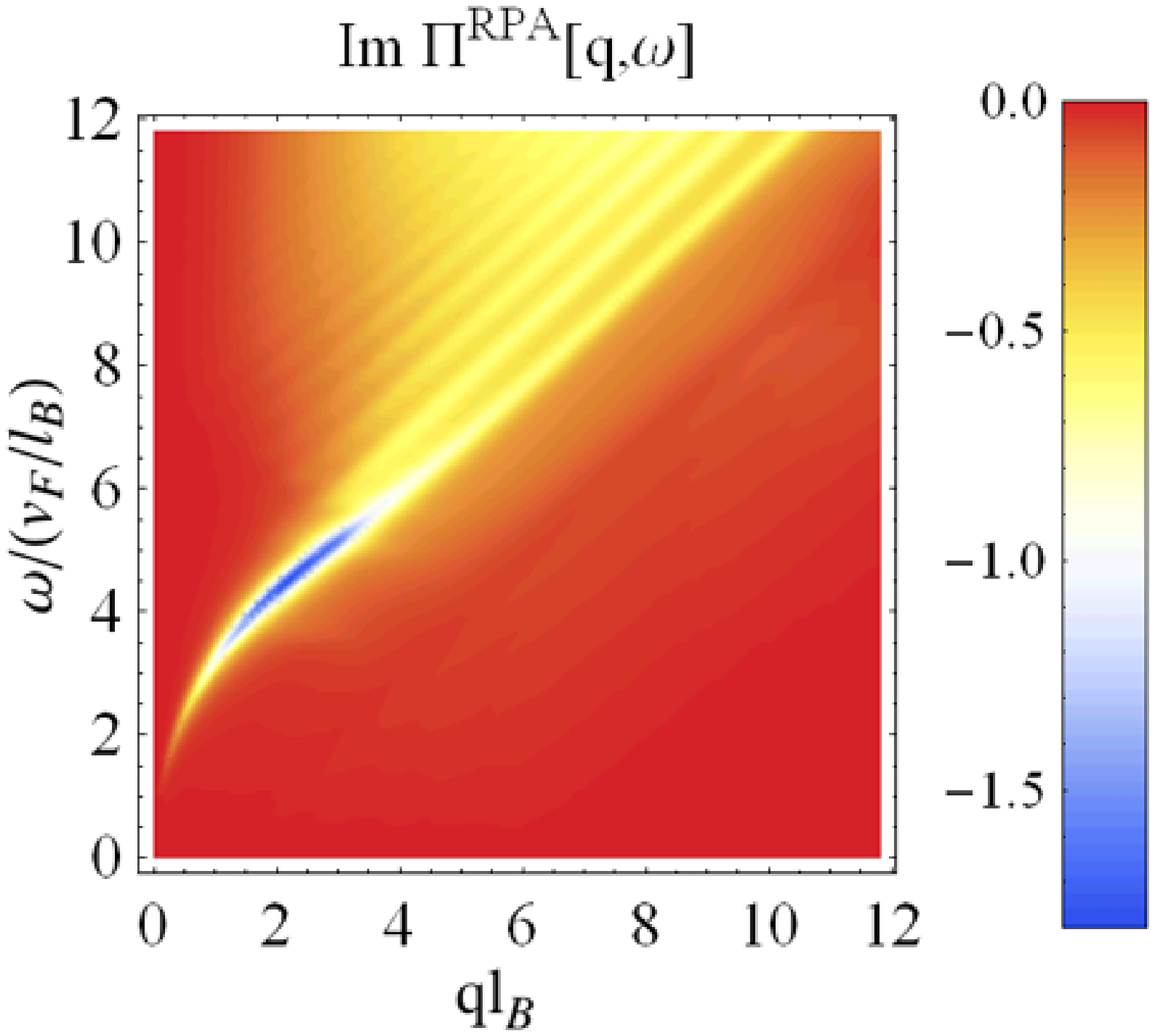}}
  \subfigure[]{\label{ImPiRPA_NF1_delt0_05}\includegraphics[width=0.36\textwidth]{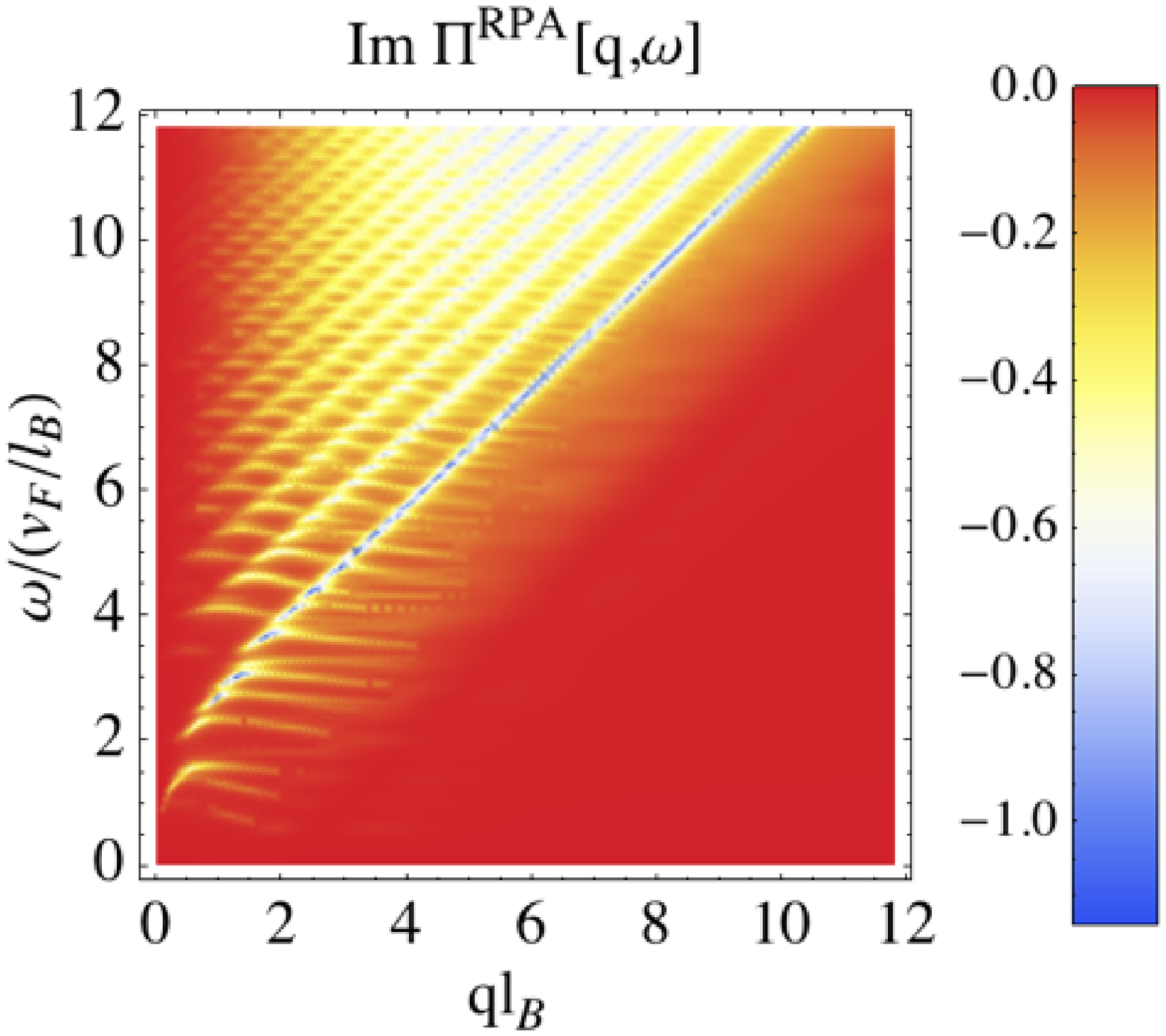}}
   \subfigure[]{\label{ImPiRPA_NF6_delt0_05}\includegraphics[width=0.36\textwidth]{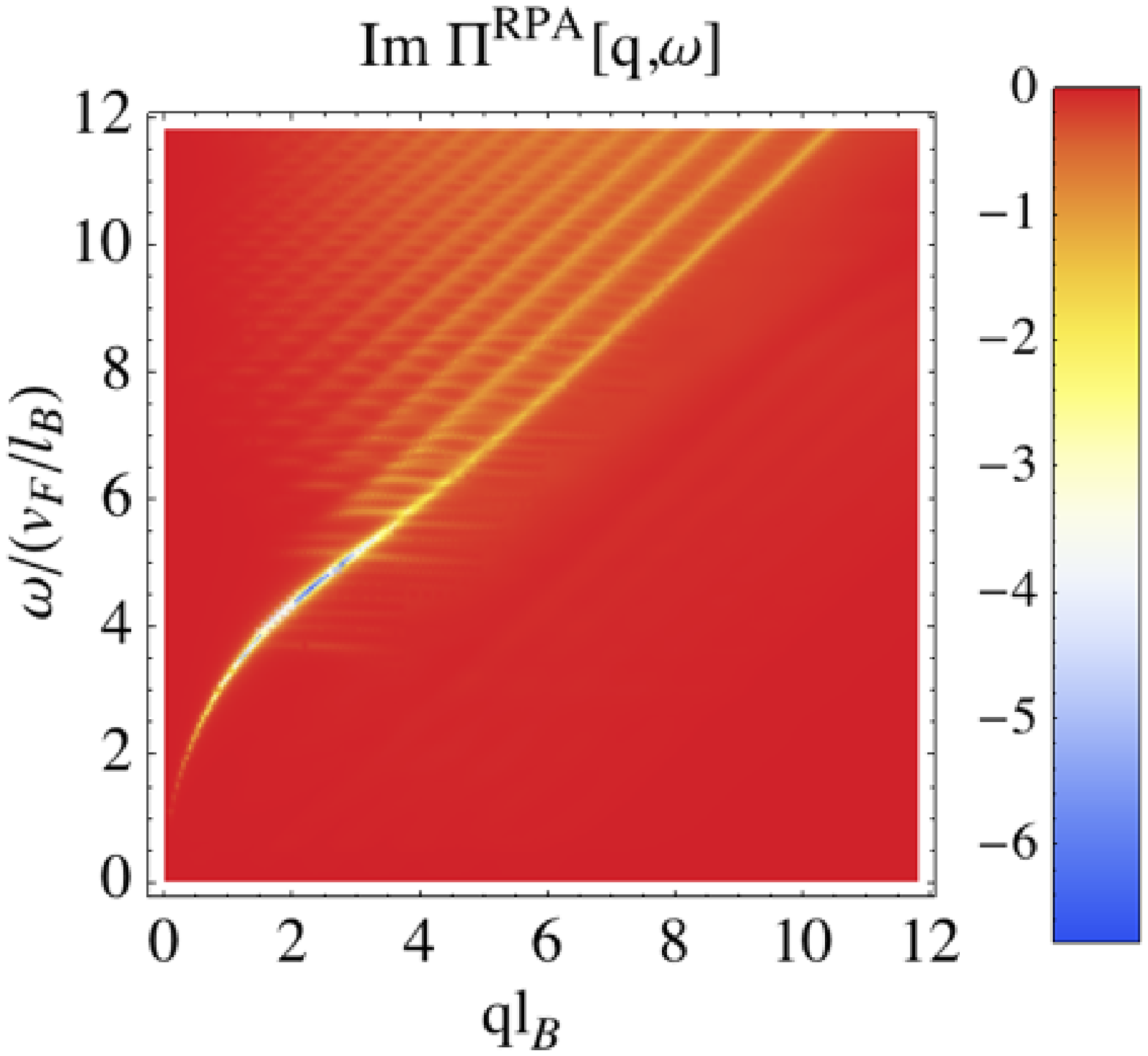}}
  \caption{\footnotesize (Color online) Density plot of ${\rm Im}\,\Pi^{RPA}(\bq,\omega)$ for graphene, for disorder broadening $\delta=0.2\vf/l_B$ [plots (a) and (b)] and $\delta=0.05\vf/l_B$ [plots (c) and (d)], and doping $N_F=1$ in plots (a) and (c), and $N_F=6$ in plots (b) and (d).}
  \label{PHESNFdelt}
\end{figure*}

The structure of the PHES in graphene and the 2DEG may be understood in the following manner. First, one notices that the boundaries
[black lines in Fig. \ref{PiDinamic}] are due to the momentum conservation of the electron-hole pairs. 
The boundaries reflect the different $B=0$ behavior of the dispersion relations in graphene (linear in $q$) as compared to the 2DEG
(quadratic in $q$). Second, for $B\neq 0$, the zero-field continuum becomes chopped into horizontal lines. The main
difference between the PHES in the 2DEG and that in graphene stems from the different LL quantization. 

In the case of the 2DEG, one
notices a constant spacing of the horizontal lines, $\omega=m\omega_c$. 
There are several contributions to each horizontal line from the allowed LL transitions, $n\rightarrow n'$, with 
$m=n'-n$. For a given integer $m$, the number of contributing LL transitions is the minimal value $\min\{N_F+1,m\}$
of $N_F+1$ and $m$. The lowest-energy horizontal line ($m=1$), thus, arises from a single LL transition, $N_F\rightarrow N_F+1$. In this
case, the line consists of $N_F+1$ well-separated {\sl islands} [see arrows in Figs. \ref{ImPi0DP2DEG} and 
\ref{ImPi02DEGzoom_delt0_05}], which is the name we give to regions of high spectral weight, and which are separated by regions of low spectral weight reflecting the nodes of the LL wave-functions. For larger values of $m$, several LL transitions $n\rightarrow n'$ contribute to the same horizontal line. Each of these transitions has $n+1$ islands. Because this number is different for each transition, the islands that arise from different transitions overlap and fill in the region of low spectral weight. As a result, the horizontal lines appear continuous [compare the $m=1$ and $m=3$ horizontal line in Figs. \ref{ImPi0DP2DEG} and \ref{ImPi02DEGzoom_delt0_05}] and the node structure of the LL wave-function is visible only in form of faint shadows parallel to the boundaries of the PHES [see Fig. \ref{ImPi0DP2DEG} and \ref{ImPiRPADP2DEG}].

In graphene, the {\it relativistic} LL quantization gives rise to horizontal lines at energies 
\begin{equation}\label{Eq:MEGraphene}
\omega=\vf l_B^{-1}[\sqrt{2n'}-\lambda \sqrt{2n}], 
\end{equation}
where we have assumed that the Fermi energy lies in the conduction band, an therefore $\lambda'=1$.
In contrast to the 2DEG, with equally spaced LLs, only a single transition $\lambda,n\rightarrow n'$ contributes to 
each horizontal line. As a consequence, the node structure of the LL wavefunctions, and therefore that of the form factor (\ref{FF}), is clearly visible in every horizontal line and not only in the one of lowest energy [see Fig. \ref{ImPi0zoom_delt0_05}]. This explains the presence of $n+1$ islands in the horizontal line corresponding to the $n\rightarrow n'$ transition. These nodes give rise to well-defined diagonal
dark lines of low spectral weight parallel to the boundaries of the particle-hole continuum. Another consequence of the 
relativistic LL structure is that the density of horizontal lines increases with energy, contrary to the case of the 2DEG, where all
lines are separated by the constant cyclotron energy [compare Figs. \ref{ImPi0zoom_delt0_05} 
and \ref{ImPi02DEGzoom_delt0_05}]. This clear separation of islands at any energy, together with the rather large number of horizontal lines and their finite disorder-induced width, leads to a stacking of the islands of different energies. This is the origin of the appearance of diagonal regions of strong spectral weight [see Figs.\ref{ImPi0DP} and \ref{ImPi0zoom_delt0_05}]. The overlap of islands of different energies is stronger for higher values of disorder in the sample (increased LL broadening $\delta$) as well as for higher filling factors. The relative energy separation between the horizontal lines will be smaller for larger $N_F$, because in the single-particle graphene spectrum, the density of LLs increases with energy. 
Electron-electron interactions turn these original regions of strong spectral weight into coherent excitations (the upper-hybrid mode and the linear magneto-plasmons). In fact, we have studied the PHES of graphene for different values of doping (i. e. $N_F\ne 3$) and disorder broadening $\delta$ and found that the physical picture exposed above is unaltered, [see Fig. \ref{PHESNFdelt}]. As a general rule, disorder and doping favor the emergence of linear magneto-plasmons in graphene as compared to the horizontal magneto-excitions. For the experimentally relevant values of $\delta$ and $N_F$, linear magneto-plasmons therefore dominate the PHES of graphene in the integer quantum Hall regime. It is worth noticing that, at fixed carrier density, increasing the filling factor ($\sim N_F$) is equivalent to effectively lowering the magnetic field. Although Fig. \ref{PHESNFdelt} represents the PHES at fixed values of $v_F/l_B$, i. e. fixed magnetic field, one easily sees that at larger values of $N_F$, most of the spectral weight is concentrated in the upper-hybrid mode [see Fig. \ref{ImPiRPA_NF6_delt0_05}]. This mode is the only collective excitation in the integer quantum Hall regime that evolves continuously into the plasmon mode at zero field. In contrast to the upper-hybrid mode, the linear magneto-plasmons become less intense in the large-$N_F$ limit and evolve for $B\rightarrow 0$ into the incoherent particle-hole continuum.

\section{Conclusions}

In conclusion, the magnetic-field particle-hole excitation spectrum in
graphene has been investigated, and the results have been compared to those of a standard 2DEG with a parabolic band dispersion. Most saliently, the particular LL quantization in graphene yields linear magneto-plasmon modes, which are not captured in the usual magneto-exciton approximation. This is due to the fact that a single particle-hole process contributes to a given energy in graphene, whereas for a 2DEG, there are, in general, many processes  contributing to the same energy. As a consequence, the node structure of the LL wavefunctions in graphene is evident at any energy of the PHES, leading to a particular spectrum where the relevant modes are diagonal, dispersing roughly parallel to $\vf q$. The dominant role of linear magneto-plasmons is enhanced by doping and Landau level broadening due to disorder. In addition to linear magneto-plasmons, electron-electron interactions give rise to the upper-hybrid mode as well, which is the plasmon mode renormalized by the magnetic field. This alternative scenario of linear magneto-plasmons, as opposed to horizontal magneto-excitons, may find an experimental proof in the framework of microwave absorption or inelastic (Raman) light scattering, which has proven to be a powerful tool for measuring the collective excitations of the 2DEG in a strong magnetic field, see e. g. Ref. \onlinecite{EW99}.

\begin{acknowledgments}
We thank Y. Gallais, P. Lederer, and F. Pi\'echon for useful
discussions and acknowledge funding from the ANR under Grant No. ANR-06-NANO-019-03 and ``Triangle de la Physique''.
\end{acknowledgments}

\bibliography{BibliogrGrafeno}

\end{document}